\documentclass[11pt]{llncs}

\usepackage{graphicx}
\usepackage{amsmath}

\newcommand{\be}{\begin{equation}}
\newcommand{\ee}{\end{equation}}
\newcommand{\ba}{\begin{eqnarray}}
\newcommand{\ea}{\end{eqnarray}}
\newcommand{\nn}{\nonumber \\}

\begin{document}

 \title{Adiabatic quantum optimization with qudits}
 \author{M.~H.~S.~Amin \and
  Neil G. Dickson \and
  Peter Smith}
 \institute{D-Wave Systems Inc., 100-4401 Still Creek Drive, Burnaby, B.C., V5C 6G9, Canada}
\maketitle

\begin{abstract}

Most realistic solid state devices considered as qubits are not true two-state
systems but multi-level systems. They can approximately be considered as qubits only
if the energy separation of the upper energy levels from the lowest two is very
large. If this condition is not met, the upper states may affect the evolution and
therefore cannot be neglected. Here, we consider devices with double-well potential
as basic logical elements, and study the effect of higher energy levels, beyond the
lowest two, on adiabatic quantum optimization. We show that the extra levels can be
modeled by adding additional (ancilla) qubits coupled to the original (logical)
qubits. The presence of these levels is shown to have no effect on the final ground
state. We also study their influence on the minimum gap for a set of 8-qubit spin
glass instances.
\end{abstract}

\section{Introduction}

Quantum information processing has been one of the fastest growing interdisciplinary
research areas over the past decade, with the promise of revolutionizing the concept
and possibilities of future computation \cite{Nielsen}. Unlike classical information
processing in which information is stored as classical states of the classical bits,
in quantum information processing, information is stored as quantum states of qubits
which are two-state quantum systems representing the basic logical elements of
quantum information. Most realistic devices, however, are not ideal two-state
systems. Therefore, the ideal two-state (qubit) model, commonly used in quantum
information, is only an approximation to their true quantum behavior. For example,
all superconducting qubits, whether charge \cite{ChargeQubits}, hybrid
\cite{HybridQubits}, phase \cite{PhaseQubits}, or flux
\cite{Mooij,Grajcar,qubitpaper,Castellano} qubits, have several energy levels, only
lowest two considered as the relevant qubit states.

\begin{figure}[t]
    \includegraphics[width=7.1cm]{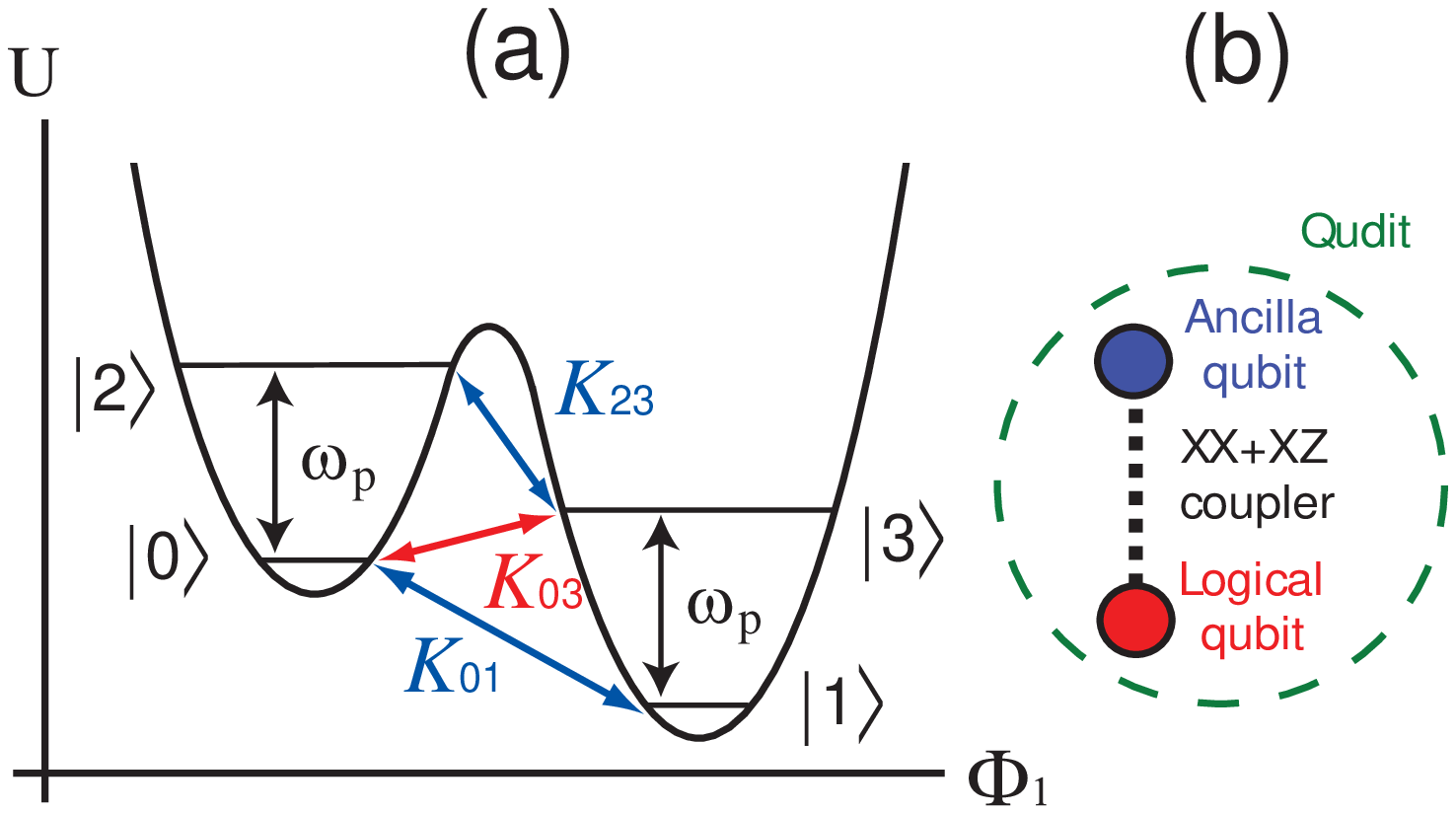}
    \includegraphics[width=5.5cm]{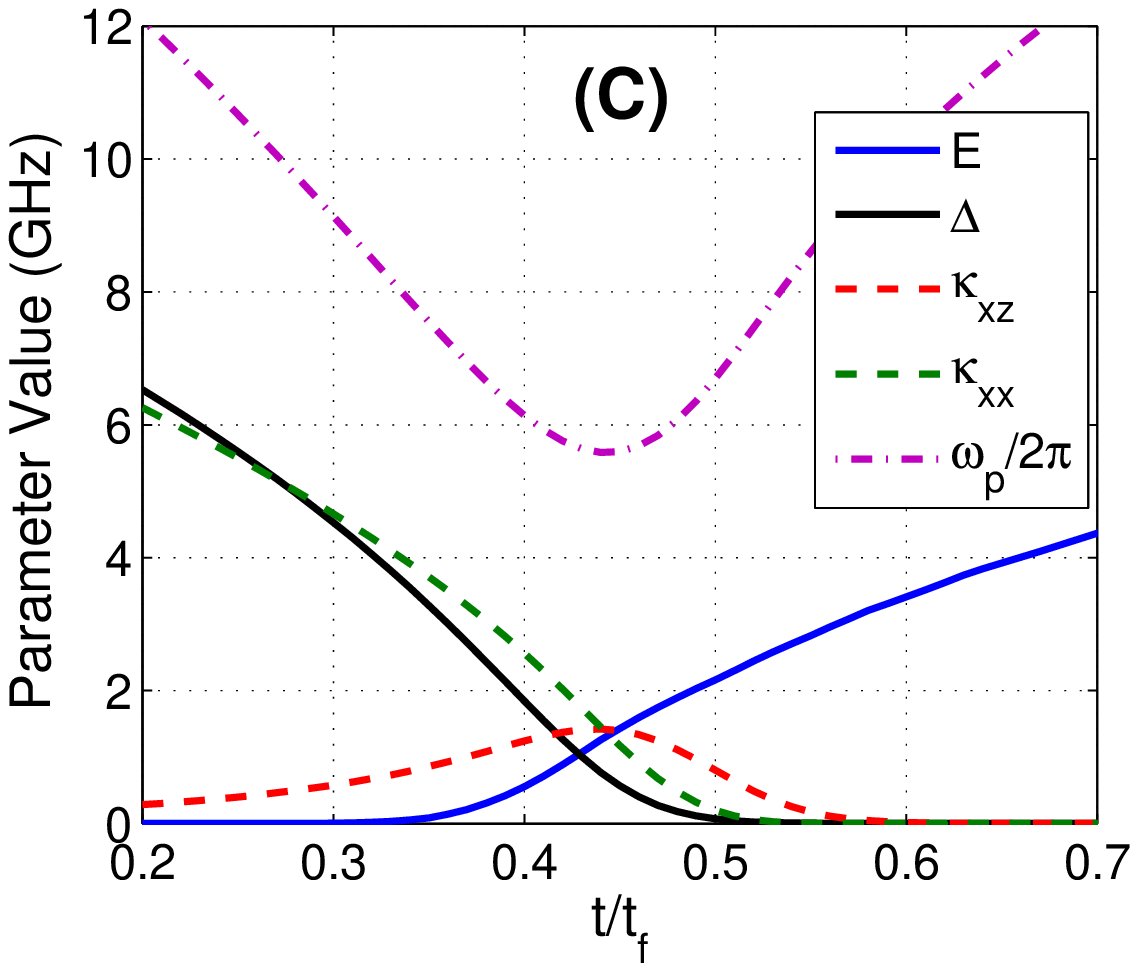}
\caption{(a) Schematic diagram of a double-well potential with the first four
metastable energy levels and the tunneling amplitudes ($K_{ij}$) between them. (b) A
four-state model of the system represented by two coupled qubits. (c) Parameters of
the qudit Hamiltonian (\ref{Hqudit}) during an annealing process, calculated for an
rf-SQUID with experimentally motivated parameters (see the appendix). All energy
scales are divided by $h$ (Planck constant) and converted into GHz.}
    \label{fig1}
\end{figure}

Flux qubits generally have potential energy (in the flux basis) in the form of a
double-well potential similar to the one illustrated in Fig.~\ref{fig1}a
\cite{Mooij,Grajcar,qubitpaper,Castellano}. In the limit of infinite barrier height
between the two wells, one can treat each well separately and obtain two sets of
quantized energy levels localized within the wells. At finite barrier height, these
localized states are not true eigenstates of the Hamiltonian anymore and therefore
are metastable, which means a system initialized in one of them will eventually
transition out of it via a tunneling process. The true eigenstates of the Hamiltonian
are indeed coherent mixtures or superpositions of those localized states. One may
still represent the system in the basis of those localized states by introducing
off-diagonal elements of the Hamiltonian which provide tunneling amplitudes between
them.

It is assumed that when the two lowest energy states dominate the dynamics of the
system at low temperatures and small energy bias, one can consider this two-state
system as a qubit. However, if the energy bias is comparable to the energy
separations within each well, $\omega_p$, or the temperature is large enough to allow
occupation of higher energy levels, then the two-state model does not describe
correctly the quantum behavior of the device. In these situations, one must include
the occupied energy levels in the description of the quantum system being studied.
Systems with $d > 2$ energy levels are commonly referred to as qu$d$its.

%

One of the important paradigms of quantum information processing is adiabatic quantum
computation (AQC) \cite{Farhi} which is known to be a universal model of quantum
computation \cite{Aharonov,Lidar}. In AQC, the Hamiltonian of the
system, generally written as 
$
 H_S(t) = \Delta(t){\cal H}_B + E(t){\cal H}_P, 
$ 
slowly evolves as $\Delta(t)$ decreases and $E(t)$ increases monotonically with time
$t$. At the beginning of the evolution, $E(0)\approx 0$ and the Hamiltonian is
dominated by the dimensionless Hamiltonian ${\cal H}_B$, with a ground state that is
usually a superposition of all states in the computation basis. At the end of the
evolution, $\Delta(t_f)\approx 0$ and the Hamiltonian is dominated by the
(dimensionless) problem Hamiltonian ${\cal H}_P$. If the evolution is slow enough,
the final state of the system will represent the ground state of ${\cal H}_P$ with
high fidelity, which is designed to solve the intended problem. The time-dependent
energy scales $\Delta(t)$ and $E(t)$ are usually not independent and both are
controlled by one external parameter.
A typical example of these functions based on a superconducting
realization of the hardware \cite{Harris10} is provided in Fig.~\ref{fig1}.

In this article, we focus on a special version of AQC known as adiabatic quantum
optimization (AQO). In AQO, ${\cal H}_P$ is diagonal in the logical basis, therefore
the final ground state is a classical state that minimizes the energy, hence can be
considered as an optimized solution to a cost function. In the literature, AQO is
sometimes called quantum annealing \cite{Brooke,Santoro}, as it uses quantum
fluctuations for annealing in a similar way as thermal fluctuations are used in
classical annealing. We are mainly interested in the transverse Ising Hamiltonian:
 \ba
 {\cal H}_B = -{1\over2}\sum_{i=1}^N \sigma_x^{(i)}, \qquad
 {\cal H}_P = \sum_{i=1}^Nh_i\sigma_z^{(i)} + \sum_{i,j=1}^{N}
 J_{ij}\sigma_z^{(i)}\sigma_z^{(j)}. \label{HP}
 \ea
where, $\sigma_{x,z}^{(i)}$ are Pauli matrices corresponding to the $i$th qubit, and
$h_i$ and $J_{ij}$ are dimensionless energy biases and coupling coefficients
respectively. An interesting and important question is, how would an adiabatic
quantum optimizer perform using realistic multi-level devices (qudits) instead of
idealized qubits.

The paper is organized in the following way. In Sec.~2, we consider a quantum device
with a double-well potential, such as a superconducting flux qubit, as a physical
implementation of a qubit. We describe such a system by an effective tunneling
Hamiltonian with finite number of levels and represent that with a few coupled
qubits. We derive the parameters of the coupled qubit Hamiltonian representing the
qudit in terms of the original tunneling Hamiltonian. In Sec.~3 we discuss how AQO is
possible with such multi-qudit system. We study the effect of the extra energy levels
on the minimum gap. Section 4 summarizes our findings and provides conclusions to our
results.

In the main body of this paper we treat the single logical element of quantum
information as a double-well potential illustrated in Fig.~\ref{fig1}. The main
discussion of the paper is independent of the physical structure behind the
double-well potential. We provide a detailed discussion of an example, i.e.,
rf-SQUID, in the appendix. We use such an rf-SQUID Hamiltonian with realistic
parameters to find the parameters of the qudit Hamiltonian used in our numerical
simulations. All parameters are extracted from experimental implementation, therefore
the numerical results we provide here are expected to closely represent the physical
reality.

\section{Single qudit Hamiltonian}


Let us consider a system with a double-well potential similar to the one depicted in
Fig.~\ref{fig1}a. As we discussed before, we would like to write the Hamiltonian of
this system in the basis of states that are localized within the wells. Such states
are not true eigenfunctions of the Hamiltonian and therefore are metastable towards
tunneling to the opposite well. Therefore, the resulting Hamiltonian in this basis
will have off-diagonal terms corresponding to transitions between states in opposite
wells. However, there are no off-diagonal terms corresponding to transitions within a
single well, as we require states within a well to be stationary. Intra-well
transitions are induced only by environmental relaxation.

Let $|l\rangle$ denote localized states within the wells. We use
even (odd) state numbers, i.e., $l=2n$ ($2n{+}1$), with
$n=0,1,2,...$, to denote states that are localized in the left
(right) well (see Fig.~\ref{fig1}a). For the lowest $M$ energy
levels ($M$ is taken to be even), the effective $M {\times} M$
tunneling Hamiltonian is written as
 \ba
 H_S = \sum_{l=0}^{M-1} E_l|l\rangle\langle l| +
 \sum_{n,m=0}^{M/2-1} K_{2n,2m{+}1}
 (|2n\rangle\langle 2m{+}1|+|2m{+}1\rangle\langle 2n|) \label{Htunneling}
 \ea
where $E_l$ is the energy expectation value for state $|l\rangle$ and $K_{2n,2m{+}1}$
is the tunneling amplitude between states $|2n\rangle$ and $|2m{+}1\rangle$, which
exist in the opposite wells. Notice that there is no matrix element between states on
the same well: $\langle 2n|H_S|2m\rangle = \langle 2n{+}1|H_S|2m{+}1\rangle = 0$,
which means that the states are metastable only towards tunneling to the other side,
or the states are quasi-eigenstates of the Hamiltonian within their own sides. In the
appendix, we explain how to arrive at such an effective tunneling Hamiltonian for a
bistable rf-SQUID and how to extract the parameters of such Hamiltonian from the
original rf-SQUID Hamiltonian.

Reading out the state of the device is usually done by identifying left and right
wells, hence logical ``0" and ``1" states. For example, in a flux qubit the two
directions of the flux generated by the persistent current in the superconducting
loop identify the two wells of the potential, hence the logical ``0" and ``1" states.
One therefore measures the magnetic flux at the end of the evolution to detect the
logical state of the system. Since all energy levels in the left (right) well yield a
negative (positive) flux, we associate all of them with logical ``0" (``1").
Therefore, all the levels within one well are {\em logically equivalent}. In other
words, the quantum numbers distinguishing states within a well are logically
irrelevant. Those degrees of freedom, however, may participate in the dynamics and
have to be taken into account in the quantum dynamics when studying the performance
of such a system.


In principle, there could be a large number of energy levels in the
full spectrum of the double-well potential, not all of them relevant
for the dynamics. Let us assume that there are a total of $M=2^m$
relevant states that participate in the dynamics. We can denote
state $|l\rangle$ by $|x_{m-1}... \,x_2 x_0\rangle$, with
$x_i\in\{0,1\}$, where $l = \sum_{i=0}^{m-1} 2^ix_i$, hence the
string $x_{m-1}... \,x_2 x_0$ is the binary representation of $l$.
With the above even-odd representation of states, all states on the
left well correspond to $x_0=0$ (even binary numbers), and all
states on the right well correspond to $x_0=1$ (odd binary numbers).
Now assume that each variable $x_i$ represents the state of one
qubit. Thus, we have an effective system of $m$ qubits representing
the qudit. Only one of the qubits, i.e., the one representing $x_0$,
determines the logical state of the system. We therefore call that
qubit ``logical'', and the other $m{-}1$ ones ``ancilla'' qubits.

For simplicity we focus only on $M=4$ levels (i.e., $m=2$ qubits representing a
qudit). We denote the left states as $|00\rangle$ and $|10\rangle$ and the right
states as $|01\rangle$ and $|11\rangle$. These four states can be represented by two
coupled qubits, as shown in Fig.~\ref{fig1}b, the bottom (top) one is taken to be the
logical (ancilla) qubit. In order to distinguish between logical and ancilla qubits
in the Hamiltonian, we label the Pauli matrices associated with the logical qubit by
$\sigma_\alpha$ and those associated with the ancilla qubit by $\tau_\alpha$, where
$\alpha = x, z$. We use the convention $\sigma_z|0\rangle = -|0\rangle$, and
$\sigma_z|1\rangle = |1\rangle$, and similarly for $\tau_z$. The effective two qubit
Hamiltonian can therefore be written as
 \ba
 H_{eff} = -{1\over 2}(\epsilon\sigma_z + \Delta\sigma_x)
 + {1\over 2}[\omega_p\tau_z + \kappa_{xz} \sigma_x(1+\tau_z) + \kappa_{xx}
 \sigma_x\tau_x ]. \label{Hqudit}
 \ea
State $|0\rangle$ of the ancilla qubit corresponds to the two lowest energy states in
the two wells and the $|1\rangle$ of the ancilla qubit provides the two upper energy
levels, which are separated from the lower ones by an energy difference equal to the
plasma frequency $\omega_p$ ($\hbar=1$). It is easy to show that (\ref{Hqudit}) is
equivalent, up to a constant energy, to Hamiltonian (\ref{Htunneling}), with $M=4$,
if
 \be
 \epsilon = E_0 - E_1 = E_2 - E_3, \qquad
 \omega_p = E_2 - E_0 = E_3 - E_1,
 \ee
 \be
 \Delta = -2K_{01}, \qquad
 \kappa_{xz} = K_{23}-K_{01} \approx K_{23}, \qquad
 \kappa_{xx} = 2K_{03} = 2K_{12}.
 \ee
As can be seen, the coupling between logical and ancilla qubits are of XX+XZ type.
Figure \ref{fig1}c illustrates typical values of these parameters during an annealing
process based on the experimentally realized 8-qubit processor described in
Ref.~\cite{Harris10} (see the appendix).

\section{Adiabatic quantum optimization with qudits}

We now generalize the above formulation for a single qudit to a coupled multi-qudit
system. The coupling is usually via the dominant degree of freedom, which for flux
qubits is the magnetic flux. In the multi-qubit representation of a qudit discussed
in the last section, the state of the logical qubits represent the direction of the
flux degree of freedom. Therefore the coupling should be via the logical qubits,
leading to a ZZ coupling term in the Hamiltonian. In practice, the magnetic flux
generated by the flux qubit is different if the qubits is in its excited state within
a well. As a result, the effect of the ancilla qubit states on the overall qubit
coupling is small, and thus is neglected in our formulation.

\begin{figure}[t]
    \centering
    \includegraphics[width=7.5cm]{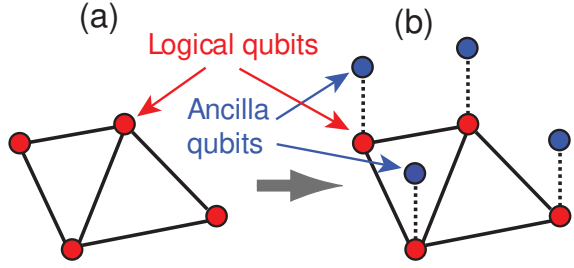}
\caption{(a) An example of a multi-qubit graph consisting 4 qubits (circles) and 5 ZZ
couplers (solid lines). (b) The same graph made of 4 qudits, each being a four-state
system represented by a logical (bottom) and an ancilla (top) qubit, coupled to each
other via a XX+XZ coupler (dashed lines).}
    \label{fig2}
\end{figure}

Given a graph of coupled qubits, in order to recast the problem in the form of
coupled qudits, it is enough to connect to each qubit in the original problem an
ancilla qubit via an XX+XZ coupling. Ancilla qubits are coupled only to their
corresponding logical qubits; there is no coupling between the logical and ancilla
qubits of other qudits. Figure \ref{fig2} illustrates such a situation for an example
graph of four coupled qubits.

In the case of $N$ qubits, the corresponding Hamiltonian is
 \ba
 && H_S = \Delta(t){\cal H}_B + E(t){\cal H}_P \label{Mqudit} \\
 &&{+}{1\over 2}\sum_{i=1}^N \left[ \omega_p^{(i)}(t)
 \tau_z^{(i)} {+} \kappa_{xz}^{(i)}(t)
 \sigma_x^{(i)}(1{+}\tau_z^{(i)}) {+} \kappa_{xx}^{(i)}(t) \sigma_x^{(i)}\tau_x^{(i)}
 \right]. \nonumber
 \ea
where ${\cal H}_B$ and ${\cal H}_P$ are dimensionless Hamiltonians defined by
(\ref{HP}). The first line is identical to the ordinary multi-qubit transverse Ising
Hamiltonian and the second line is the contribution of the ancilla qubits.

Figure \ref{fig1}c shows the result of calculation of parameters of Hamiltonian
(\ref{Hqudit}) as a function of time during the annealing process, using an rf-SQUID
model with experimentally motivated parameters as explained in the appendix. As
expected $\Delta(t)$ decreases and $E(t)$ increases monotonically with time as
required. Notice the non-monotonic variation of $\omega_p$, which represents the
energy separation of the upper energy levels from the lowest two (see
Fig.~\ref{fig1}a). To be able to approximate a qudit by a two-state qubit, it is
necessary that the excited states within each well be far above the lowest energy
levels. This puts a limit on the parameter of the problem Hamiltonian: $h_i,J_{ij}
\ll \omega_p/E(t)$. If this condition is not met, the energy bias of the qubits can
become so large that the lowest state in one well becomes close to the excited state
in the next well (see Fig.~\ref{fig1}a as an example), hence the two-state model for
the device fails to hold. The important question is: Can we still solve the intended
optimization problems using such devices? As we shall see the answer is yes.

At the end of the evolution, the barrier between the two wells becomes so large that
all tunnelings between the lowest energy levels will stop. This means that all the
off-diagonal elements of the Hamiltonian will vanish: $\Delta = \kappa_{xx}^{(i)} =
\kappa_{xz}^{(i)} =0$. It is easy to see that in the absence of these off-diagonal
elements, the logical and ancilla qubits decouple from each other and the ground
state of the total Hamiltonian (\ref{Mqudit}) will be the ground state of the
original Ising Hamiltonian ${\cal H}_P$ for the logical qubits with all the ancilla
qubits being in state $|0\rangle$. Therefore, for the purpose of optimization, the
ancilla qubits have no effect on the final ground state even if $h_i,J_{ij} >
\omega_p/E(t)$. They, however, affect the dynamics of the system as we shall see.
This means that {\em AQO can be performed using qudits instead of qubits regardless
of the final Hamiltonian}. The effect of the upper energy levels on the computation
time is yet to be discussed.

In a closed system, the computation time $t_f$ is considered to be inversely related
to the size of the minimum gap $g_{\rm min}$ between the ground state and the first
excited state. This gap is commonly obtained using the Hamiltonian of coupled ideal
qubits, i.e., the first line of (\ref{Mqudit}). Naively, one might think that adding
the ancilla qubits would increase the total number of qubits and therefore would
significantly reduce the size of the minimum gap $g_{\rm min}$. One should keep in
mind that the ancilla qubits are just added to model the upper energy levels that
already exist in the spectrum of the actual physical device that plays the role of
qubit. Therefore, the ancilla qubits should always be added if one wants to have a
correct description of the real physical system. Since the number of upper energy
levels are typically large, a large number of ancilla qubits is necessary to
accurately describe all the energy levels. This means that even a small size system
should be represented by a large number of qubits. Therefore, if the above statement
is correct all realistic systems should have extremely small gaps. Intuitively, one
expects that if the upper energy levels are far away from the lowest two levels, then
their influence should be negligible. Thus, the presence of the ancilla qubits should
not significantly affect the gap as long as $\omega_p$ is much larger than other
terms in the original Hamiltonian. As is clear from Fig.~\ref{fig1}c, $\omega_p$ is
not very large, but still almost a factor of 3 larger than $E(t)$ for the second half
of the evolution. Therefore, it is expected that the excited levels, or equivalently
the ancilla qubits, have some (but small) influence on the minimum gap and the
evolution.

\begin{figure}[t]
    \centering
    \includegraphics[width=7cm]{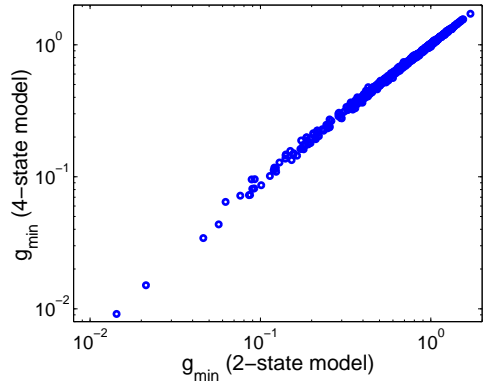}
\caption{Comparison of the minimum gap between two-state and four-state models. }
    \label{gmin}
\end{figure}

We test this conjecture by comparing the values of $g_{\rm min}$ calculated with and
without the ancilla qubits. To achieve this, we considered an ensemble of 8-qubit
spin glass instances generated with random parameters: $h$'s and $J$'s were selected
uniformly at random from $0, \pm 1/7, ..., \pm 6/7, \pm 1$, except for $J$'s that
don't represent an edge in the complete bipartite graph $K_{4,4}$, whose values were
all zero. Among 800 instances generated, 669 of them had non-degenerate ground state
for which we calculated $g_{\rm min}$ using exact diagonalization technique. Figure
\ref{gmin} shows a comparison between $g_{\rm min}$ calculated without the ancilla
qubits (referred to as two-state model) and with the ancilla qubits (referred to as
four-state model). As is clear from the figure, for some instances the size of the
gap is increased and for some is decreased. On average the change of $g_{\rm min}$
was below 1\%. For the 4 data points available in the small gap region, the size of
the gap was reduced by a maximum of 36\%. More investigations, especially on larger
problem instances, is necessary to determine if this is a trend for all small gap
instances. Nevertheless, considering doubling the number of effective qubits from
two-state model to four-state model, the above change is not significant.

%
%

\section{Conclusions}

The effect of the energy levels above the lowest two levels in the physical
implementation of a qubit on adiabatic quantum optimization was studied. We discussed
a model of a multi-level system with a double-well potential as a system of coupled
qubits. One of the qubits (logical qubit) represents the logical state of the
physical system and the remaining qubits (ancilla qubits) produce the upper energy
levels. At the end of the evolution, the solution to the problem described by
Hamiltonian $H_P$, is determined only by the state of the logical qubits. We showed
that the state of the logical qubits in the final ground state is unaffected by the
ancilla qubits, regardless of $H_P$ (as long as potential bistability is preserved).
Therefore, the same problems that can be solved by ideal qubits can also be solved
using qudits. We studied the influence of the ancilla qubits on the minimum gap and
showed that for realistic qubit parameters the effect for average problems is small.
Problems with very small gap sizes had systematically smaller gaps when ancilla
qubits were introduced into the formulation. However, the four data points in our
simulation is not sufficient to predict a trend, especially for large scale problems.
When generalized to open quantum systems using density matrix methods, the results of
this model are in close agreement with results from experiments on a system of eight
coupled rf-SQUID qubits. Discussion of the open quantum calculation and experiments
is beyond the scope of this paper and will be discussed in a future publication.

\section*{Acknowledgment}

We are thankful to  A.J. Berkley, P. Bunyk, S. Gildert, F. Hamze, R.
Harris, J. Johansson, M.W. Johnson, K. Karimi, T. Lanting, and G.
Rose for fruitful discussion, and C.J.S. Truncik for help at the
initial stages of the programming.

\newpage

\appendix

\section*{Appendix: rf-SQUID Hamiltonian}

In this appendix, we look closely at a specific example of qubit implementation,
namely a compound Josephson junction rf-SQUID \cite{qubitpaper}. This choice was
motivated by recent experimental progress in implementing multi-qubit quantum
annealing process with such qubits \cite{Harris10}. All parameters for our numerical
simulations are based on the 8-qubit unit cell studied in Ref.~\cite{Harris10}. The
qubit itself is discussed in detail in Ref.~\cite{qubitpaper}, but here we only
consider a simplified version.

\begin{figure}[h]
    \centering
    \includegraphics[width=4.5cm]{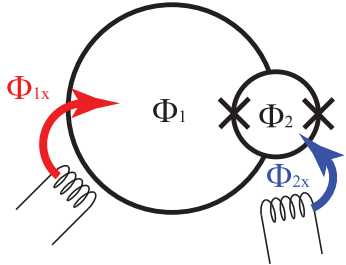}
\caption{Schematic diagram of a tunable rf-SQUID. The external fluxes $\Phi_{1x}$ and
$\Phi_{2x}$ control the energy bias and tunneling amplitude respectively.}
    \label{rfSQUID}
\end{figure}

The qubit, as illustrated in Fig.~\ref{rfSQUID}, has two main superconducting loops
and therefore two flux degrees of freedom $\Phi_1$ and $\Phi_2$, subject to external
flux biases $\Phi_{1x}$ and $\Phi_{2x}$, respectively. The Hamiltonian of the qubit
is written as
 \ba
 H_{\rm SQUID}
 = {q_1^2 \over 2C_1} + {q_2^2 \over 2C_2} + U(\Phi_1,\Phi_2), \label{HS}
 \ea
where $C_1$ and $C_2$ are parallel and series combinations of the junction
capacitances, $q_1$ and $q_2$ are the sum and difference of the charges stored in the
two Josephson junctions respectively, and
 \ba
 U(\Phi_1,\Phi_2) =  {(\Phi_1-\Phi_{1x})^2/2L_1}
 +  {(\Phi_2-\Phi_{2x})^2/2L_2} \nn
 - 2E_J \cos (\pi \Phi_2/\Phi_0)\cos(2\pi \Phi_1/\Phi_0)
 \qquad \label{U}
 \ea
is a 2-dimensional potential with $L_i$ being the inductance of the $i$th loop and
$\Phi_0 {=}\, h/2e$ is the flux quantum. We have assumed symmetric Josephson
junctions with Josephson energies $E_J {=}\, I_c\Phi_0/2\pi$, where $I_c$ is the
junctions' critical current. (A small asymmetry can be tuned away in situ in the
physical implementation \cite{qubitpaper}.)

At $\Phi_{1x}\approx \Phi_0/2$, the potential can become bistable and therefore form
a two-dimensional double-well potential. If $L_2$ is small enough so that the
deviation of $\Phi_2$ from $\Phi_{2x}$ can be neglected, then the two-dimensional
classical potential $U(\Phi_1,\Phi_2)$ can be approximated by a one-dimensional
double-well potential, as shown in Fig.~\ref{fig1}a. However, with our realistic
qubit parameters, $\Phi_2$ cannot be neglected and therefore is accounted for in all
our numerical calculations. When $\Phi_{1x} = \Phi_0/2$, the two wells are symmetric
with no energy bias ($\epsilon=0$). One can tilt the potential by changing
$\Phi_{1x}$ and establish an energy bias $\epsilon$, as depicted in Fig.~\ref{fig1}a.
It is also possible to change the barrier height by changing $\Phi_{2x}$. Quantum
annealing is performed by slowly increasing the barrier height from a very small
value to a very large value through ramping $\Phi_{2x}$. Details of the annealing
process and techniques used to make all terms in the Hamiltonian change uniformly are
discussed in Ref.~\cite{Harris10}. At the end, the system behaves as the Hamiltonian
(\ref{Mqudit}) with all its time-dependent parameters determined experimentally. Our
goal here is to extract these parameters numerically for the rf-SQUID Hamiltonian
(\ref{HS}) having known all parameters ($L_i, C_i, I_{c}$) and the experimental
values of the external fluxes $\Phi_{1x}(t)$ and $\Phi_{2x}(t)$ as a function of
time.


The eigenvalues $E_n$ and eigenstates $|E_n\rangle$ of the rf-SQUID Hamiltonian
(\ref{HS}) can be calculated by numerical diagonalization. They, however, are not
directly useful for simulations of AQO in a multi-qubit system defined by Hamiltonian
(\ref{Mqudit}). One therefore needs to derive (\ref{Mqudit}), or single qudit version
of it (\ref{Htunneling}), from those eigenvalues and eigenstates. In principle, it is
possible to write down the Hamiltonian in a basis defined by states $|l\rangle$
localized in the wells, instead of the energy basis $|E_n\rangle$, as long as they
form (at least approximately) an orthonormal basis. Our numerical procedure is as
follows. First, we numerically diagonalize the original rf-SQUID Hamiltonian
(\ref{HS}) to obtain energy eigenstates $|E_n\rangle$. We then select the first $M$
eigenstates and diagonalize the flux operator $\Phi_1$ in such subspace. This way we
find $M$ flux states $|\chi_i\rangle$ with eigenvalues $\chi_i$ each being a
superposition of states $|E_n\rangle$. Some of the flux states will have negative and
some positive induced flux $\delta \Phi_i=\chi_i-\Phi_{1x}$. We treat states with
negative (positive) induced flux as states localized in the left (right) well. We
then separate these two set of localized states and once again diagonalize the
rf-SQUID Hamiltonian (\ref{HS}), but now separately in each left and right subspaces.
The final result is a Hamiltonian that looks like (\ref{Htunneling}). Different
matrix elements of the resulting Hamiltonian determine different terms in
(\ref{Htunneling}) which in turn determine the parameters of the qudit Hamiltonian
(\ref{Hqudit}).

\end{document}